\documentclass{article}

\usepackage[preprint,nonatbib]{neurips_2022}

\usepackage[utf8]{inputenc} 
\usepackage[T1]{fontenc}    
\usepackage{hyperref}       
\usepackage{url}            
\usepackage{booktabs}       
\usepackage{amsfonts}       
\usepackage{nicefrac}       
\usepackage{microtype}      
\usepackage{xcolor}         
\usepackage{graphicx}
\usepackage{wrapfig}
\hypersetup{colorlinks,allcolors=blue!50!black}

\usepackage{pmboxdraw}
\usepackage{newunicodechar}
\newunicodechar{└}{\textSFii}
\newunicodechar{├}{\textSFviii}
\newunicodechar{─}{\textSFx}
\newunicodechar{│}{\textSFxi}

\usepackage[
    backend=biber,
    style=apa,
    natbib=true,
    doi=false,
    isbn=false,
    url=false,
]{biblatex}
\title{Sorrel: A simple and flexible framework for multi-agent reinforcement learning}

\author{%
  Rebekah A. Gelpí$^{1,2,3}$\thanks{$^1$ Department of Psychology, University of Toronto \\
  \hspace*{0.49cm} $^2$ Schwartz Reisman Institute for Technology \& Society, University of Toronto \\
  \hspace*{0.49cm} $^3$ Vector Institute \\
  \hspace*{0.49cm} $^4$ Department of Computer Science, University of Toronto} \\
  \texttt{rebekah.gelpi@mail.utoronto.ca} \\
  \And
  Yibing Ju$^4$\hspace*{1cm} \\
  \texttt{bing.ju@mail.utoronto.ca}\hspace*{1cm} \\
  \And
  Ethan C. Jackson$^{1,3}$ \\
  \texttt{jackson.ethan.c@gmail.com} \\
  \And
  Yikai Tang$^{1,3}$ \\
  \texttt{yikai.tang@mail.utoronto.ca} \\
  \And
  Shon Verch$^{4}$ \\
  \texttt{verchshon@gmail.com} \\
  \And
  Claas Voelcker$^{3,4}$ \\
  \texttt{cvoelcker@cs.utoronto.edu} \\
  \And
  William A. Cunningham$^{1,2,3,4}$ \\
  \texttt{wil.cunningham@utoronto.ca} \\
}

\begin{document}

\maketitle

\begin{abstract}
  We introduce \textbf{Sorrel} (\href{https://github.com/social-ai-uoft/sorrel}{github.com/social-ai-uoft/sorrel}), a simple Python interface for generating and testing new multi-agent reinforcement learning environments. This interface places a high degree of emphasis on simplicity and accessibility, and uses a more psychologically intuitive structure for the basic agent-environment loop, making it a useful tool for social scientists to investigate how learning and social interaction leads to the development and change of group dynamics. In this short paper, we outline the basic design philosophy and features of Sorrel.
\end{abstract}

\section{Introduction}

Understanding how individuals act, learn, and change within social systems forms a core challenge for researchers across the social sciences. These disciplines have used varying lenses to provide insight, ranging from studies of individuals' capacities for reasoning with and about others, to macro-level investigations of collective behaviour.

Historically, a major challenge to modelling these processes has been their complexity. Mathematical models in the fields of game theory, economics, and cultural evolution  \parencite[e.g.][]{blume2018population, boyd1985culture, boyd2009voting, cavalli1981cultural, lewis2012transmission, schelling1971dynamic} have successfully captured many social phenomena such as cultural transmission, tensions between individual and collective benefits, and inequality between social groups. However, these models have been traditionally limited in their ability to express individuals' learning, representational systems, and complex choice ecology.

One promising route to addressing these limitations has been the use of the multi-agent reinforcement learning (MARL) approach to model social systems. Although these models can be significantly more demanding in terms of the resources they require, advances in machine-learning algorithms and computational efficiency have made it possible to represent dynamic social systems in much greater detail and fidelity. MARL models have provided compelling evidence of how interactions between individuals can produce complex and varied social dynamics \parencite[for reviews, see][]{hertz2025beyond, wong2023deep}, including the emergence of social norms \parencite{koster2022spurious, vinitsky2023learning, tang2023unequal}, cooperation \parencite{DBLP:journals/corr/LererP17, oroojlooy2023review, du2023review}, communication systems \parencite{lipowska2022emergence, karten2023role, eccles2019biases}, social roles \parencite{wang2020roma, wang2021rode}, and stereotypes \parencite{gelpi2025social, duenez2021statistical}.

While the MARL framework is powerful and has seen increasing application by social scientists, many existing tools for generating and testing MARL environments are relatively complex, making them more difficult to use for researchers without extensive programming knowledge. To remedy this problem, we introduce \textbf{Sorrel} (SOcial Recreations for REinforcement Learning), a Python interface for developing new MARL environments which places an emphasis on simplicity, flexibility, and accessibility for social scientists, and which allows for the modelling of various emergent phenomena among large groups of learning agents.

In the following sections, we briefly introduce the design philosophy and features of Sorrel, and highlight options available for use and extension.

\section{Background and Related Work}

With the increasing popularity of MARL as a method not only for understanding dynamic social behaviour but also other dilemmas such as logistics, cybersecurity, and optimization problems, several APIs exist that allow for testing models within multiagent environments. We briefly summarize some recent implementations:

\paragraph{PettingZoo} Extending the popular Gym/Gymnasium interfaces for single-agent reinforcement learning \parencite{brockman2016openai, towers2024gymnasium} to a MARL setting, PettingZoo \parencite{terry2021pettingzoo} includes both a suite of pre-existing environments and a Gym-like grammar for developing new environments.

\paragraph{Melting Pot} The Melting Pot framework \parencite{agapiou2022melting, leibo2021scalable} provides a comprehensive suite of environments that feature desirable characteristics for modelling a variety of scenarios and properties of social interactions, such as free-rider problems, the presence of mixed motives, and coordination problems. The environments are diverse and allow for testing many different models and hypotheses, but support for extending the codebase to build new environments is limited by the need for familiarity with both Python and Lua.

\paragraph{JaxMARL} The JaxMARL library \parencite{rutherford2024jaxmarl} adapts several MARL benchmarks and introduces two new environments for use with models written using the Jax library for deep RL \parencite{jax2018github}. This allows for faster training and evaluation than previous approaches. No explicit methods for extending the 9 prebuilt environments exist.

Our goal in designing Sorrel is to make the process of designing new environments for testing models within MARL settings accessible and approachable for social scientists. Thus, our interface uses an intuitive nested structure in which individual agents and entities exist within an environment. These agents, in turn, perceive the environment using an embedded observation system, evaluate these observations using an embedded model, and act using an embedded action system.

\section{Features}

Currently, Sorrel's focus is on providing an interface for extensible gridworld games. Thus, it includes several features for: creating objects within an environment that either behave stochastically, or dynamically change over time or in response to agents' actions; embedding agents within these environments; giving these agents varying forms of perception in the environment; using various model implementations to drive agents' action policies; and visualizing the trajectories of agents and environments over time.

\subsection*{Environment}

All elements necessary for training agents are encapsulated within an \textit{environment}. This includes the world within which agents exist and interact, methods for setting up the environment, and a method for running experiments within the environment. Sorrel has the following nested structure for environment:


\begin{verbatim}
    Environment
    ├─ World
    │  ├─ World map
    │  │  ├─ Entities
    │  │  │  ├─ Entity properties
    │  │  │  └─ Entity transition function
    │  │  └─ Agents
    │  │     ├─ Agent properties
    │  │     ├─ Observation specification
    │  │     ├─ Model specification
    │  │     ├─ Action specification
    │  │     └─ Agent transition function
    │  ├─ World properties
    │  └─ World functions
    │     └─ Helper functions: add(), remove(), observe()...
    ├─ Environment properties
    ├─ Environment transition function
    └─ Environment functions
       ├─ Setup: setup_agents(), populate_world()...
       └─ Experiment runner: run_experiment()
\end{verbatim}

\paragraph{World} Environments contain a \textit{world} which includes a \textit{world map} and a set of helper functions for placing, moving, and removing all entities that exist within the environment at any given time. Any time an agent acts within the world, it does so by interacting with entities (non-agentic objects such as walls, trees, or food) or other agents embedded within the world map.

The world in Sorrel environments changes as a product of the environment's \textit{transition function}. This function steps through changes that occur in the world due to agents' actions, stochastic processes, or predictable dynamics.

\paragraph{Entity transitions} Even though entities do not act upon the world, they can still change over time: a food resource can grow over time or decline due to harvesting by an agent, water can become polluted due to lack of cleaning or be cleaned by an agent, and certain objects might randomly spawn or despawn from an environment. 

\paragraph{Agent transitions} The central reinforcement learning loop is implemented within agents' transition function. Within this loop, agents obtain their \textbf{state} from an observation function, get an \textbf{action} from the model and action specifications, obtain their \textbf{reward}, and receive information regarding whether the agent is \textbf{done} after acting on the environment.

\subsection*{Agent Architecture}

We anticipate that agents' transition functions will contain the information of primary importance and interest for social scientists, as agents' states, actions, and rewards will represent the relationship between what agents can perceive and the outcomes of this perception on behaviour and learning. Agents with differing capacities for perception might act in very different ways and obtain very different rewards; similarly, agents using different models might differ in their actions and reward given the same observations, and agents with different available actions could also differ even with the same input and model architecture. We therefore place an emphasis on modularity and flexibility in agents' designs, to facilitate the implementation of agents that vary along all three of these dimensions.

\paragraph{Observation specifications}

Depending on the environment and the goals of a researcher, it may be desirable for an agent to have full knowledge of an environment, or only be able to partially observe its immediate vicinity. Similarly, agents could observe the environment as an image if using a CNN architecture, as a matrix of one-hot coded entities if using a DQN or similar, or as a series of ASCII strings if being passed to an LLM.

In addition to these pre-existing implementations, the base observation class can be flexibly extended to allow for arbitrary observation formats. For example, observations could include not only an observation of the environment, but also other environment features or internal properties of the agent (e.g., an agent's inventory).

\paragraph{Model and action specifications}

Models take an input in the form of an observation, and yield an output in the form of an action. They thus link the agents' observation specification, which dictates the format of agents' observations, with the agents' action specifications, which dictate how many actions an agent has and how they map onto an agent's possible actions.

We include a pre-built model implementing a combination of the IQN \parencite{dabney2018implicit} and Rainbow \parencite{hessel2018rainbow} models, written in PyTorch. We also include a model class that allows a human player to participate interactively in an environment, either as a single agent, or with other models whose policies are governed by an algorithm. In addition to this, we include a simple model wrapper that allows researchers to bring their own models, as long as these models implement a small number of shared model functions that prompt the model to act on the environment and (optionally) train the model.

\section{Default environments}

\begin{wrapfigure}[15]{r}{0.5\textwidth}
    \centering
    \vspace{-1.5em}
    \includegraphics[width=\linewidth]{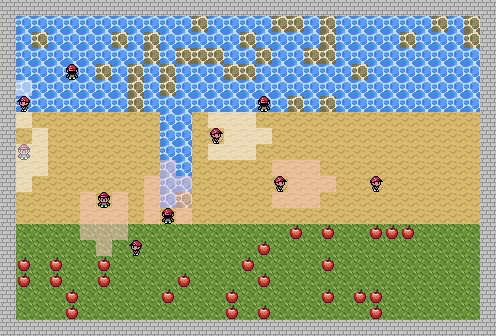}
    \caption{Visualization of the \textbf{Cleanup} environment \parencite[adapted from][]{agapiou2022melting} implemented within Sorrel.}
    \label{fig:cleanup}    
\end{wrapfigure}

Our initial release includes two environments. The first is a basic environment known as Treasure Hunt, which we use as a tutorial environment that covers the basics of designing environments. In this environment, agents compete to obtain valuable gems that occasionally appear in the environment. We also include an implementation of the Cleanup environment \parencite{agapiou2022melting} from Melting Pot.

Both environments include a set of default sprites that allow these environments to use the included visualization utilities. These utilities allow researchers to animate multiple turns in an environment and observe agents' behaviour unfold over time.

\section{Documentation and Tutorials}

The \href{https://sorrel.readthedocs.io/en/latest/}{documentation} for Sorrel includes references for the base environment classes, model classes, and visualization tools, as well as several tutorials that walk through the process of building an environment from scratch and using models with Sorrel.

\printbibliography






\end{document}